\documentclass[pre,nofootinbib,amsmath,amssymb,preprint,showpacs]{revtex4}
\usepackage[T1]{fontenc} \usepackage[latin1]{inputenc}
\usepackage{amsmath} \usepackage{graphicx} \usepackage{amssymb}

\begin{document}

\title{Rouse Modes of Self-avoiding Flexible Polymers}

\author{Debabrata Panja} \affiliation{Institute for Theoretical
Physics, Universiteit van Amsterdam, Valckenierstraat 65, 1018 XE
Amsterdam, The Netherlands} \author{Gerard
T. Barkema$^{\dagger\ddagger}$} \affiliation{$^{\dagger} $Institute
for Theoretical Physics, Universiteit Utrecht, Leuvenlaan 4, 3584 CE
Utrecht, The Netherlands\\ $^{\ddagger}$Instituut-Lorentz,
Universiteit Leiden, Niels Bohrweg 2, 2333 CA Leiden, The Netherlands}

\begin{abstract} 
Using a lattice-based Monte Carlo code for simulating self-avoiding
flexible polymers in three dimensions in the absence of explicit
hydrodynamics, we study their Rouse modes.  For self-avoiding
polymers, the Rouse modes are not expected to be statistically
independent; nevertheless, we demonstrate that numerically these modes
maintain a high degree of statistical independence.  Based on
high-precision simulation data we put forward an approximate
analytical expression for the mode amplitude correlation functions for
long polymers. From this, we derive analytically and confirm
numerically several scaling properties for self-avoiding
flexible polymers, such as (i) the real-space end-to-end distance,
(ii) the end-to-end vector correlation function, (iii) the correlation
function of the small spatial vector connecting two nearby monomers at
the middle of a polymer, and (iv) the anomalous dynamics of the middle
monomer. Importantly, expanding on our recent work on the theory of
polymer translocation, we also demonstrate that the anomalous dynamics
of the middle monomer can be obtained from the forces it experiences,
by the use of the fluctuation-dissipation theorem.
\end{abstract}


\maketitle

\section{Introduction\label{sec1}}

Polymer dynamics is a field where first-principle analytical
derivations of most quantities of interest from microscopic monomeric
movements are difficult to come by. Indeed, the answer to the question
why phantom polymers (theoretical realizations of polymers that can
intersect themselves) remain, to this day, the central pillar for
studying polymer dynamics is easily traced to the fact that they allow
full analytical calculations of essentially all of their dynamical
properties. In the absence of explicit hydrodynamics, a phantom
polymer's dynamics is described by the so-called Rouse equation that
forms the basis of all analytical calculations \cite{doi,de}. The Rouse
equation holds in the high-viscosity limit of the surrounding medium,
and the dynamics of the monomers are ``overdamped'', i.e., the
velocity of each monomer at any given time is proportional to the
total force it experiences. In the Rouse equation the total force in
any monomer is comprised of the spring forces due to its neighboring
monomers, and the random thermal forces from the surrounding
medium. The linearity of the Rouse equation allows one to decompose it
into linear dynamical equations of independently evolving (Rouse)
modes, which are the Fourier transforms of the monomer co-ordinates in
three dimensional space.  In detail, we consider a polymer of length
$N$, consisting of $N+1$ monomers connected sequentially by $N$ bonds
(harmonic springs). The positions $\vec{R}_n(t)$ of monomers
$n=0,\ldots,N$ at time $t$ can then undergo Fourier transformation,
yielding for the  amplitude of the $p$-th mode ($p$ is an integer
$\ge0$)
\begin{eqnarray}
\vec{X}_p(t)=\frac{1}{N+1}\sum_{n=0}^N\cos\left[\frac{\pi(n+1/2)p}{N+1}\right]\,
\vec{R}_n(t).
\label{e1}
\end{eqnarray}
The cornerstone of all analytical calculations for phantom polymer
dynamics is the following relation for $p,q\neq0$, derived exactly
from the Rouse equation:
\begin{eqnarray}
X_{pq}(t)=\langle\vec{X}_p(t)\cdot\vec{X}_q(0)\rangle\propto\frac{N}{p^2}\,\exp\left[-A\,\frac{p^2}{N^2}\,t\right]\,\delta_{pq},
\label{e2}
\end{eqnarray}
where $\delta_{pq}$ is the Kronecker delta function, and $A$ a
constant. All throughout this paper, the angular brackets represent an
average over the equilibrium ensemble of polymers.  Equation
(\ref{e2}) is further supplemented by $X_{0p}(t)=0$ for $p\neq0$, and
$X_{00}(t)=m\delta(t)$, where $\vec{X}_0(t)$ is the location of the
center-of-mass of the polymer at time $t$, and $m\propto1/N$ is the
mobility of the center-of-mass of the polymer. The Kronecker delta
terms signify the statistical independence of the modes of a phantom
polymer.  Using these mode amplitude correlation functions, the
quantities of interest for a phantom polymer can be analytically
tracked by reconstructing them from the modes \cite{doi,de}. However, in
reality, a polymer is not phantom, but is self-avoiding. The
self-avoidance property introduces long-range correlations along the
backbone of the polymer, and it destroys the linearity of the Rouse
equation, making a similar [to Eq. (\ref{e2})] equation for the mode
amplitude correlation functions impossible to derive from the
appropriately formulated Rouse equation. Because of this reason, it
comes as no surprise to us that for self-avoiding polymers we have not
been able to find, in published literature, a comprehensive study of
the mode amplitude correlations that connects to the scaling
properties of self-avoiding polymers.

The purpose of this paper is to put forward an {\it approximate}
analytical expression for the mode amplitude correlation functions for
long polymers, which is then corroborated with extensive simulations
of a lattice-based model for the dynamics of self-avoiding flexible
polymers in three dimensions in the absence of explicit hydrodynamics.
To be more precise, with $\nu\approx0.588$ (the Flory exponent in
three dimensions), and $A_1$ and $A_2$ two constants, we demonstrate
that for a self-avoiding Rouse polymer of length $N$
\begin{eqnarray}
X_{pq}(t)=\langle\vec{X}_p(t)\cdot\vec{X}_q(0)\rangle\approx
A_1\,\frac{N^{2\nu}}{p^{1+2\nu}}\,\exp\left[-A_2\,\frac{p^{1+2\nu}}{N^{1+2\nu}}\,t\right]\,\delta_{pq}
\label{e3}
\end{eqnarray}
for $p,q\neq0$ holds up to a very good approximation for long
polymers, while $X_{0p}(t)=0$ for $p\neq0$ and
$X_{00}(t)\propto(1/N)\delta(t)$ holds exactly. The reasons why Eq.
(\ref{e3}) is not exact are the following: (a) although numerically
the modes maintain a high degree of statistical independence, the mode
amplitude correlation functions are not exactly statistically
independent (we do not expect to be so anyway);  and (b) for higher
modes ($p,q\gtrsim6$) there are small deviations from exponential
behavior at long times. Nevertheless, given that $X_{pp}(0)$ is a
rapidly decaying function of $p$, the dominant contribution of the
modes to quantities of interest for the polymer at long time-scales
come from the lower modes. As a result, such quantities can be
analytically reconstructed from Eq.~(\ref{e3}). We demonstrate this by
using Eq.~(\ref{e3}) to derive several scaling properties
for self-avoiding polymers, such as (i) the real-space end-to-end
distance, (ii) the end-to-end vector correlation function, (iii) the
correlation function of the small spatial vector connecting two nearby
monomers at the middle of the polymer, and (iv) the anomalous dynamics
of the middle monomer. Some of these scaling laws can also be obtained
from the dynamic scaling law~\cite{doi}. Note that except the case
for characterizing the anomalous dynamics of the middle monomer, all
the other quantities (as above) concern only the polymer's internal
structure; implying that for (i-iii) we only need Eq. (\ref{e3}),
while for the anomalous dynamics of the middle monomer we also need
$X_{00}(t)$ and $X_{0p}(t)$ for $p\neq0$.  Note also that the standard
Rouse result Eq.~(\ref{e2}) is recovered when we simply replace $\nu$
by $1/2$ for a phantom polymer. Apparently, the dominant consequence
of volume exclusion is the different size scaling of the self-avoiding
chains. The inability for polymers to cross each other does not seem to
cause large cross-correlations between different modes.  Thus, given
the content of this paper, we expect that Eq.~(\ref{e3}) will provide
researchers a way forward for analytical treatment of the properties of
self-avoiding polymers.

The structure of this paper is as follows. In Sec. \ref{sec2} we
describe our polymer model and demonstrate the scaling property
Eq.~(\ref{e3}). In Sec. \ref{sec3} we use Eq.~(\ref{e3}) to obtain the
scaling properties for the real-space end-to-end distance and the
end-to-end vector correlation function. In Sec. \ref{sec4} we derive
the scaling properties of the small spatial vector connecting two
nearby monomers. In Sec. \ref{sec5} we derive the anomalous dynamics
of the middle monomer, and by expanding our recent work on the theory
of polymer translocation, show that the anomalous dynamics of the
middle monomer can also be obtained by using the
fluctuation-dissipation theorem. Our conclusions are then summarized
in Sec. \ref{sec6}.

\section{Our polymer model and the scaling properties of
  the mode amplitude correlation functions\label{sec2}}

\subsection{Our polymer model and the calculation of the mode
  amplitudes\label{sec2a}}

Over the last years, we have developed a highly efficient simulation
approach to polymer dynamics. This is made possible via a lattice
polymer model, based on Rubinstein's repton model \cite{rubinstein}
for a single reptating polymer, with the addition of sideways moves
(Rouse dynamics). A detailed description of this model, its
computationally efficient implementation and a study of some of its
properties and applications can be found in
Refs. \cite{heukelum03,LatMCmodel}.

In this model, each polymer is represented by a sequential string of
monomers, living on a face-centered-cubic lattice with periodic
boundary conditions in all three spatial directions. Monomers adjacent
in the string are located either in the same, or in neighboring
lattice sites. The polymers are self-avoiding: multiple occupation of
lattice sites is not allowed, except for a set of adjacent
monomers. The polymers move through a sequence of random
single-monomer hops to neighboring lattice sites. These hops can be
along the contour of the polymer, thus explicitly providing reptation
dynamics. They can also change the contour ``sideways'', providing
Rouse dynamics. Each kind of movement is attempted with a statistical rate
of unity, which provides us with the definition of time. This model has
been used before to simulate the diffusion and exchange of polymers in
an equilibrated layer of adsorbed
polymers~\cite{klein_adsorbed}. Recently, we have used this code
extensively to study polymer translocation under a variety of
circumstances \cite{wolterink06,trans1,trans2,trans3,trans3a,vocks08},
and also to study the dynamics of polymer adsorption \cite{adsorb}.

Given that our model has periodic boundary conditions in all three
spatial directions, we use the following definition for the mode
amplitude:
\begin{eqnarray}
\vec{X}_p(t)=\frac1{N+1}\sum_{n=0}^N[\vec{R}_n(t)-\vec{R}_0(t)]\cos\left[\frac{\pi(n+1/2)p}{N+1}\right],
\label{e4}
\end{eqnarray}
where $\vec{R}_0(t)$ is the location of the zero-th monomer at time
$t$. In this definition we only need the spanning vectors, i.e.,
monomer co-ordinates w.r.t. that of monomer zero. These spanning
vectors are obtained from a summation over the bond vectors between
monomers adjacent in the string. This completely avoids the invocation
of periodic images, and allows for spanning distances which exceed
half the box size.  Note that Eq. (\ref{e4}) can be derived from
Eq. (\ref{e1}) as follows. First we use
\begin{eqnarray}
\frac1{N+1}\sum_{n=0}^N\cos\left[\frac{\pi(n+1/2)p}{N+1}\right]\,\cos\left[\frac{\pi(n+1/2)p'}{N+1}\right]=\frac12\,\delta_{p,p'},
\label{e5}
\end{eqnarray}
in order to express $\vec{R}_n(t)$, the location of the $n$-th monomer
w.r.t. that of the center-of-mass
$\vec{R}_{cm}(t)\equiv\vec{X}_{0}(t)$, from the inverse Fourier
transform of Eq. (\ref{e1}), as
\begin{eqnarray}
\vec{R}_n(t)=\vec{R}_{cm}(t)+2\sum_{p=1}^\infty\vec{X}_p(t)\,\cos\left[\frac{\pi(n+1/2)p}{N+1}\right].
\label{e6}
\end{eqnarray}
Then we write
\begin{eqnarray}
\vec{R}_0(t)=\vec{R}_{cm}(t)+2\sum_{p=1}^\infty\vec{X}_p(t)\,\cos\left[\frac{\pi
p}{2(N+1)}\right],
\label{e7}
\end{eqnarray}
i.e.,
\begin{eqnarray}
\vec{R}_n(t)-\vec{R}_0(t)=2\sum_{p=1}^\infty\vec{X}_p(t)\,\left\{\cos\left[\frac{\pi(n+1/2)p}{N+1}\right]-\cos\left[\frac{\pi
p}{2(N+1)}\right]\right\},
\label{e8}
\end{eqnarray}
from which Eq.~(\ref{e4}) follows with the use of
Eq.~(\ref{e5}). Equation~(\ref{e4}) shows that the modes $q\neq0$
relate only to the polymers' structural configuration, as remarked
earlier.

\subsection{The scaling properties of the mode amplitude correlation
  functions\label{sec2b}}

We start with the behavior of $X_{00}(t)$. We note that the internal
forces over the entire polymer at any time sum to zero, and therefore,
(for the overdamped dynamics in the Rouse model) the motion of its
center-of-mass is simply proportional to the average thermal
fluctuating force on the entire polymer. This force is
$\delta$-correlated in time, which implies that the center-of-mass of
the polymer simply performs a random walk. In fact,
$X_{00}(t)=m'\delta(t)$, and the mobility $m'\propto1/N$. This leads
us to the result that $X_{00}(t)\propto(1/N)\delta(t)$. We will return
to the motion of the center-of-mass in Sec. \ref{sec5}.
\begin{figure}[h]
\begin{center}
\includegraphics[width=\linewidth]{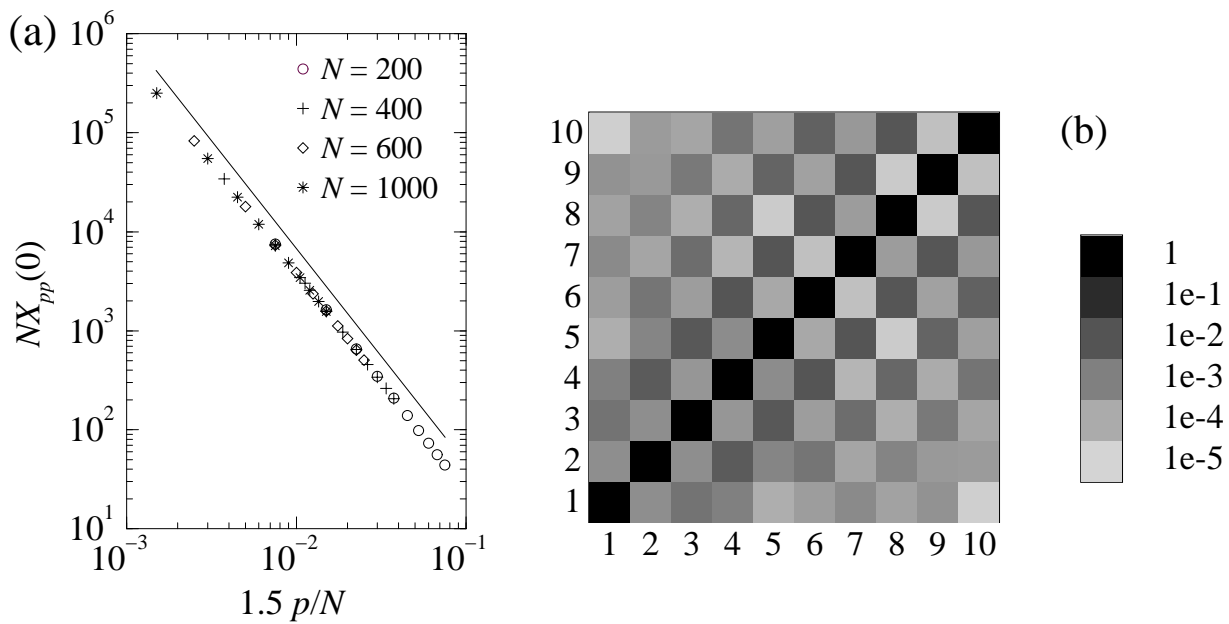}
\end{center}
\caption{(a) The scaling $X_{pp}(0)\sim N^{2\nu}/p^{1+2\nu}$
demonstrated for $p=1,\ldots,10$ for $N=200$, 400, 600 and 1000. The
solid line corresponds to
$NX_{pp}(0)\sim(N/p)^{-(1+2\nu)}\approx(N/p)^{-2.176}$. (b) The matrix
$\chi_{pq}(0)\equiv X_{pq}(0)/\sqrt{X_{pp}(0)C_{qq}(0)}$ in
logarithmic grayscale for $p,q=1,\ldots,10$ for $N=1000$ in a square
checkerboard plot. See text for details.\label{fig1}}
\end{figure}

Next, for the behavior of $X_{0p}(t)$ for $p\neq0$ we proceed as
follows. We note that $X_{0p}(0)=0$. Indeed, $X_{0p}(0)$ is the
equilibrium average over the dot product of $\vec{X}_0$ and
$\vec{X}_p$ at zero time difference, for which, for any given value of
$\vec{X}_p$, the center-of-mass is equally likely to
be present at any location within the periodic box. For the
calculation of $X_{0p}(t)$, having summed over all the locations of
$\vec{X}_0$ for a given value of $\vec{X}_p$ at the first step, and
then having summed over the configurations $\vec{X}_p$ at the second
step, we obtain $X_{0p}(0)=0$ (from the first step). Thereafter, we
use the result that the center-of-mass performs a simple random walk,
further implying that $X_{0p}(t)=X_{0p}(0)$, which equals zero.
\begin{figure}[h]
\begin{center}
\includegraphics[angle=270,width=0.6\linewidth]{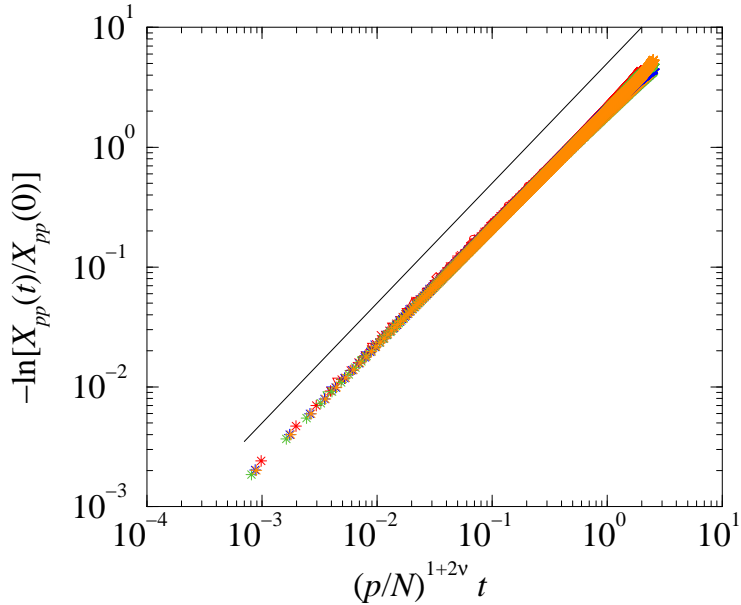}
\end{center}
\caption{Data collapse for $-\ln[X_{pp}(t)/X_{pp}(0)]$ as a function
of $(p/N)^{1+2\nu}t$, for $p=1$ (stars), 2 (triangles down), 3
(crosses), 4 (triangles up), 5 (diamonds), 6 (pluses), 7 (circles),
and $N=200$ (red), 400 (blue), 600 (green) and 1000 (orange). The
collapsed data are fitted well with an exponential; however, we note
that there are small (but systematic) deviations from the exponential
behavior for $p>>1$. The solid (black) line corresponds to
$-\ln[X_{pp}(t)/X_{pp}(0)]\sim(p/N)^{1+2\nu}t$.\label{fig2}}
\end{figure}

Finally, we demonstrate the scaling property of $X_{pq}(t)$ for
$p,q\neq0$ in Figs. \ref{fig1} and \ref{fig2}. In Fig. \ref{fig1}(a)
we present the data for $p=1,\ldots,10$ for $N=200$, 400, 600 and
1000, and demonstrate that $X_{pp}(0)\sim N^{2\nu}/p^{1+2\nu}$ up to a
high precision. In Fig. \ref{fig1}(b) we plot the matrix
$\chi_{pq}(0)\equiv X_{pq}(0)/\sqrt{X_{pp}(0)X_{qq}(0)}$ in
logarithmic grayscale for $p,q=1,\ldots,10$ for $N=1000$, wherein the
diagonal elements are unity by construction. The off-diagonal elements
of $\chi_{pq}(0)$ are typically three or more orders of magnitude
smaller than the diagonal ones.  Cross-correlations between two even
or between two odd modes are strictly {\it not\/} zero.
Cross-correlations between even and odd modes are much smaller, and it
is likely that these are also strictly {\it not\/} zero, but we cannot
ascertain that within our numerical precision. We emphasize that the
off-diagonal elements of $\chi_{pq}(0)$ not being zero is not caused
by the lack of numerical precision, or is due to artifacts of our
model --- similar features have been found in Ref. \cite{downey} for
single polymers, and in Ref. \cite{binder} for polymer melts; the
modes are simply not statistically independent for self-avoiding
polymers. Nevertheless, the precision with which the modes remain
statistically independent is remarkable.

In Fig. \ref{fig2} we plot $-\ln[X_{pp}(t)/X_{pp}(0)]$ as a function
of $(p/N)^{1+2\nu}t$ for $p=1,\ldots,7$, and $N=200$, 400, 600 and
1000, and obtain a data collapse. The collapsed data are well fitted
by an exponential behavior of $X_{pp}(t)/X_{pp}(0)$; however, we note
that there are small (but systematic) deviations from the exponential
behavior for $p\gtrsim6$. These deviations are larger for larger $p$,
although this fact is not very clearly discernible in Fig. \ref{fig2}.

As remarked before Eq.~(\ref{e3}), Figs.~\ref{fig1} and \ref{fig2}
collectively demonstrate that the scaling behavior Eq.~(\ref{e3}) is
not exact; instead, it is a rather good approximation. In the
following sections we will use the approximate scaling behavior of
Eq.~(\ref{e3}) and demonstrate that it reproduces the scaling behavior
of several observables for self-avoiding polymers.

\section{Scaling of the end-to-end distance and the
  end-to-end vector correlation function\label{sec3}}

In order to calculate the equilibrium end-to-end distance we use
Eq.~(\ref{e8}) and write
\begin{eqnarray}
\vec{R}_N(t)-\vec{R}_0(t)=2\sum_{p=1}^\infty\vec{X}_p(t)\,\left\{\cos\left[\frac{\pi(N+1/2)p}{N+1}\right]-\cos\left[\frac{\pi
    p}{2(N+1)}\right]\right\},
\label{e9}
\end{eqnarray}
from which, using the scaling relation (\ref{e3}) we extract
\begin{eqnarray}
\langle[\vec{R}_N(t)-\vec{R}_0(t)]^2\rangle\approx16A_1N^{2\nu}\sum_{p\,\in\,
    \text{\scriptsize odd}}\frac{1}{p^{1+2\nu}}\,\sin^2\left[\frac{\pi
    N
    p}{2(N+1)}\right]\nonumber\\&&\hspace{-7.6cm}=16A_1N^{2\nu}\sum_{p\,\in\,
    \text{\scriptsize
    odd}}\frac{1}{p^{1+2\nu}}\,\left\{1-\sin^2\left[\frac{\pi
    p}{2(N+1)}\right]\right\}.
\label{e10}
\end{eqnarray}
On the r.h.s. of Eq.~(\ref{e10}) the first term
$\displaystyle{\sum_{p\,\in\,\text{\scriptsize odd}}p^{-(1+2\nu)}}$
sums up to a numerical constant, and the second term
$\displaystyle{\sum_{p\,\in\,\text{\scriptsize
odd}}p^{-(1+2\nu)}\sin^2[\pi p/2(N+1)]}$ can be converted to an
integral
\begin{eqnarray}
8A_1\int_0^{\infty}dx\,\frac{\sin^2[\pi x/2]}{x^{1+2\nu}} \nonumber
\end{eqnarray}
with an integrable singularity at $x=0$. The first term produces the
well-known Flory scaling behavior
$\langle[\vec{R}_N(t)-\vec{R}_0(t)]^2\rangle\sim N^{2\nu}$, while the
second term provides a correction of $O(1)$ to the scaling which can
be neglected in the scaling limit.

The end-to-end vector correlation function can be similarly expressed
in terms of the mode amplitude correlation functions; however, in
order to obtain its scaling behavior we do not need to do so. The
point is that at long times the dominant contribution of the
end-to-end vector comes from the mode $p=1$, and thus the approximate
scaling (\ref{e3}) confirms that the end-to-end vector correlation
function decays exponentially in time, for which the characteristic
time for the decay is given by the Rouse time $\tau_R\sim N^{1+2\nu}$.

\section{The correlation function of the small
spatial vector connecting two nearby monomers at the middle of the
polymer\label{sec4}}

Consider the small spatial vector $\vec{r}_n(t)$ from monomer
$(N-n)/2$ to monomer $(N+n)/2$ with $n\ll N$. In this section we study
the behavior of
$C_n(t)\equiv\langle\vec{r}_n(t)\cdot\vec{r}_n(0)\rangle$.

By expanding in terms of modes, we have
\begin{eqnarray}
\vec{r}_n(t)=2\sum_{p=1}^\infty\vec{X}_p(t)\,\left\{\cos\left[\frac{\pi(N+n+1)p}{2(N+1)}\right]-\cos\left[\frac{\pi(N-n+1)p}{2(N+1)}\right]\right\}\nonumber\\&&\hspace{-11.7cm}=-4\sum_{p\,\in\,
    \text{\scriptsize odd}}\vec{X}_p(t)\,\sin\left[\frac{\pi
    np}{2(N+1)}\right].
\label{e11}
\end{eqnarray}
Thereafter, using the approximate scaling (\ref{e3}) we obtain
\begin{eqnarray}
C_n(t)\approx16A_1N^{2\nu}\sum_{p\,\in\, \text{\scriptsize
    odd}}\frac{1}{p^{1+2\nu}}\,\sin^2\left[\frac{\pi n
    p}{2(N+1)}\right]\,\exp\left[-A_2\,\frac{p^{1+2\nu}}{N^{1+2\nu}}\,t\right]\nonumber\\&&\hspace{-11.2cm}\approx8A_1n^{2\nu}\int_{0}^\infty
    \frac{dx}{x^{1+2\nu}}\,\sin^2[\pi
    x/2]\,\exp[-A_2x^{1+2\nu}(t/n^{1+2\nu})].
\label{e12}
\end{eqnarray}

We did not find a closed-form expression for the integral in
Eq. (\ref{e12}); hence, in order to obtain the scaling behavior of
$\langle\vec{r}_n(t)\cdot\vec{r}_n(0)\rangle$ with time, we proceed
with an approximation in the following manner. First, we note that at
long times the dominant contribution to the integral comes from
$x\lesssim A_2(t/n^{1+2\nu})^{1/(1+2\nu)}$. This observation allows us
to write
\begin{eqnarray}
C_n(t)\approx8A_1n^{2\nu}\int_{0}^{\frac{A_2}{n}\,t^{1/(1+2\nu)}}\frac{dx}{x^{1+2\nu}}\,\sin^2[\pi
  x/2].
\label{e13}
\end{eqnarray}
Next, for such (small) values of $x$ in the integral (\ref{e13}), we
can approximate $\sin(\pi x/2)$ by $\pi x/2$, leading to
\begin{eqnarray}
C_n(t)\approx2A_1\pi^2n^{2\nu}\int_{0}^{\frac{A_2}{n}\,t^{1/(1+2\nu)}}\frac{dx}{x^{1-2\nu}}\approx2\pi^2A_1A^{2(1-\nu)}_2n^{2(2\nu-1)}\,t^{-\frac{2(1-\nu)}{1+2\nu}}.
\label{e14}
\end{eqnarray}

\begin{figure}[h]
\begin{center}
\includegraphics[angle=270,width=0.6\linewidth]{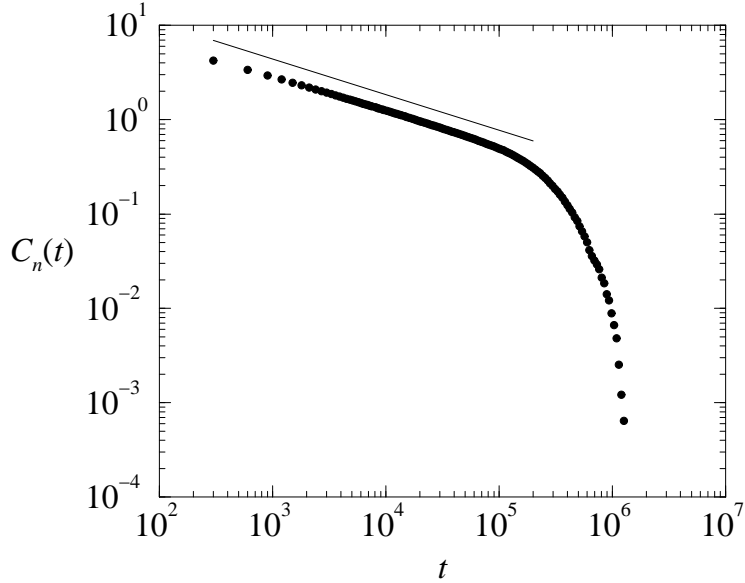}
\end{center}
\caption{Plot of $C_n(t)$ for $N=390$ and $n=6$. The solid line
corresponds to $C_n(t)\sim t^{-2(1-\nu)/(1+2\nu)}\approx
t^{-0.378}$. The steep drop in the data at very long times corresponds
to the terminal (exponential) Rouse relaxation.\label{fig3}}
\end{figure}

The behavior $C_n(t) \sim t^{-2(1-\nu)/(1+2\nu)}$ only lasts till the
Rouse time, the lifetime of the lowest mode ($p=1$) in the summation
of Eq.~(\ref{e12}). In Fig.~\ref{fig3} we numerically confirm this
behavior of $C_n(t)$ for $N=390$ and $n=6$.

\section{Anomalous dynamics of the middle monomer\label{sec5}} 

\subsection{Mean-square displacement of the middle monomer\label{sec5a}} 

To obtain the mean-square displacement of the middle monomer we first
use Eq.~(\ref{e6}) to write
\begin{eqnarray}
\vec{\Delta
    r}_{N/2}\equiv\vec{R}_{N/2}(t)-\vec{R}_{N/2}(0)=\vec{X}_{00}(t)-\vec{X}_{00}(0)+2\sum_{p\,\in\,\text{\scriptsize
    even}}[\vec{X}_p(t)-\vec{X}_p(0)].
\label{e15}
\end{eqnarray}
We then use $X_{0p}(t)=0$ and the approximate scaling relation
(\ref{e3}) to obtain
\begin{eqnarray}
\langle\Delta
    r_{N/2}^2\rangle\approx\langle[\vec{R}_{cm}(t)-\vec{R}_{cm}(0)]^2\rangle+8\!\!\sum_{p\,\in\,\text{\scriptsize
    even}}X_{pp}(0)\left[1-\frac{X_{pp}(t)}{X_{pp}(0)}\right]\nonumber\\&&\hspace{-10.1cm}
    \approx\langle[\vec{R}_{cm}(t)-\vec{R}_{cm}(0)]^2\rangle+8A_1N^{2\nu}\!\!\sum_{p\,\in\,\text{\scriptsize
    even}}\frac{1}{p^{1+2\nu}}\left\{1-\exp\left[-A_2\frac{p^{1+2\nu}}{N^{1+2\nu}}t\right]\right\}.
\label{e16}
\end{eqnarray}
Given that the center-of-mass performs a random walk, the first
(center-of-mass) term on the r.h.s. of Eq. (\ref{e16}), as argued in
Sec. \ref{sec2}, increases linearly with $t$, while in the limit
$N\to\infty$ the second term can be converted to an integral:
\begin{eqnarray}
\sum_{p\,\in\,\text{\scriptsize
    even}}\frac{1}{p^{1+2\nu}}\left\{1-\exp\left[-A_2\frac{p^{1+2\nu}}{N^{1+2\nu}}t\right]\right\}=8A_1\int_0^\infty\frac{dx}{x^{1+2\nu}}\,[1-\exp(-A_2x^{1+2\nu}t)]\nonumber\\&&\hspace{-7.3cm}\approx8A_1t^{\frac{2\nu}{1+2\nu}}\int_0^\infty\frac{dx}{x^{1+2\nu}}\,[1-\exp(-A_2x^{1+2\nu})],
\label{e17}
\end{eqnarray}
\begin{figure}[h]
\begin{center}
\includegraphics[angle=270,width=0.55\linewidth]{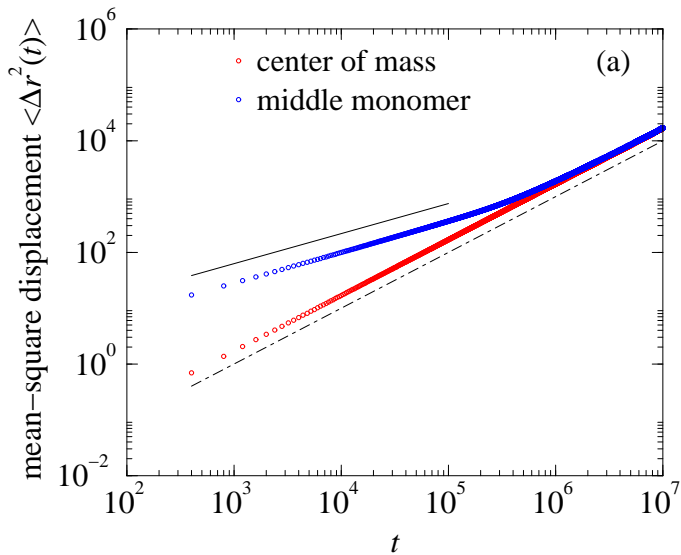}
\end{center}
\caption{The scaling behaviors for the mean-square displacement
$\langle\Delta r^2(t)\rangle$: for the center-of-mass (bottom set of
points), and for the middle monomer (top set of points). The
dot-dashed line corresponds to $\langle\Delta r^2(t)\rangle\sim t$ and
the solid line corresponds to $\langle\Delta r^2(t)\rangle\sim
t^{2\nu/(1+2\nu)}$. \label{fig4}}
\end{figure}
which, up to some constant factor, scales with time as
$t^{\frac{2\nu}{1+2\nu}}$.  Once again, the $t^{\frac{2\nu}{1+2\nu}}$
behavior of the mean-square displacement of the middle monomer will
only hold till the lifetime of the lowest mode in the summation of
Eq.~(\ref{e17}) ($p=2$) scaling as the Rouse time $\tau_R\sim
N^{1+2\nu}$, after which $\langle\Delta r_{N/2}^2\rangle$ has to
increase linearly with $t$. Note that Eqs. (\ref{e16}) and (\ref{e17})
confirm the well-known result that by the Rouse time, the middle
monomer typically displaces itself by the spatial extent of the
polymer $\sim N^{2\nu}$.

In Fig. \ref{fig4}, we confirm the above behavior of the mean-square
displacement of the middle monomer and that of the center-of-mass. We
have also checked, in support of $X_{00}(t)\sim\delta(t)$, that the
probability distribution of the displacement of the center-of-mass of
the polymer along all three spatial direction is Gaussian, with the
width scaling $\sim\sqrt{t}$ (data not shown).

\subsection{Fluctuation dissipation theorem and the anomalous dynamics
  of the middle monomer\label{sec5b}}

In Secs. \ref{sec2} and \ref{sec5} we argued that the force on the
center-of-mass of the polymer is the average over all the thermal
forces on the monomers, leading us to $X_{00}(t)\sim\delta(t)$, which
yielded the $\sim t$ behavior of the mean-square displacement of the
center-of-mass. This raises the natural question, namely: is it
possible to derive the anomalous dynamics of the middle monomer ---
i.e., the $\langle\Delta r_{N/2}^2\rangle\sim t^{\frac{2\nu}{1+2\nu}}$
up to the Rouse time and $\sim t$ thereafter --- from the combined
forces (internal and external) that act on the middle monomer?

We trace back the formulation of this problem to our recent work on
unbiased polymer translocation through a narrow pore in a membrane
\cite{trans1}. Therein we showed, in the following manner, that the
anomalous dynamics of translocation --- essentially that of the
translocating monomer --- can indeed be derived from the forces it
experiences. The velocity of translocation $v(t)$ (along the direction
perpendicular to the membrane) is related to the force $\phi(t)$ (also
acting upon it along the direction perpendicular to the membrane), via
the ``impedance'' memory kernel $\mu(t)$ and ``admittance'' memory
kernel $a(t)$, by
\begin{eqnarray}
\phi(t)=\int_0^t dt'\,\mu(t-t')\,v(t')+g(t),
\label{e18}
\end{eqnarray}
and
\begin{eqnarray}
v(t)=\int_0^t dt'\,a(t-t')\,\phi(t')+h(t).
\label{e19}
\end{eqnarray}
In Eqs. (\ref{e18}) and (\ref{e19}), $g(t)$ and $h(t)$ are the noise
terms satisfying $\langle g(t)\rangle=\langle h(t)\rangle=0$, and the
corresponding fluctuation-dissipation theorems $\langle
g(t)g(t')\rangle=\langle\phi(t)\phi(t')\rangle_{v=0}=|\mu(t-t')|$ and
$\langle h(t)h(t')\rangle=\langle
v(t)v(t')\rangle_{\phi=0}=|a(t-t')|$.
Moreover, the uniqueness of the relation
between $v(t)$ and $\phi(t)$ dictates that their Laplace transforms
must satisfy the condition $\mu(z)a(z)=1$. Finally, we
characterized the anomalous dynamics of translocation by integrating
$\langle v(t)v(t')\rangle_{\phi=0}$ twice in time: we obtained the
mean-square displacement $\langle s^2(t)\rangle$ of the translocating
monomer to behave as $t^{\frac{1+\nu}{1+2\nu}}$ until the Rouse time
$\tau_R$, beyond which $\langle s^2(t)\rangle$ linearly with $t$. We
showed that the memory kernel scheme works beautifully for
translocation out of planar confinements \cite{trans2}, translocation
by a pulling force at the head of the polymer \cite{trans3} (and
additional back-pulling voltage \cite{trans3a}), field-driven
translocation \cite{vocks08}, and polymer adsorption \cite{adsorb}.

It is easy to see in the above analysis that if
$\langle\phi(t)\phi(t')\rangle_{v=0}=|\mu(t-t')|\sim t^{-\alpha}$ for
some $\alpha$, then the mean-square displacement of the translocating
monomer has to increase as $t^\alpha$. We now show that the above
scheme established for the anomalous dynamics of polymer translocation
also works, with the modification that we now have to deal with vector
velocity $\vec{v}(t)$ of the middle monomer and vector force
$\vec{f}(t)$ on it, for the anomalous dynamics of the middle monomer
of a self-avoiding polymer; namely that we will show that
$|\mu(t-t')|\sim t^{-\frac{2\nu}{1+2\nu}}$, which would imply
$\langle\Delta r_{N/2}^2\rangle\sim t^{\frac{2\nu}{1+2\nu}}$. Of
course such power-law behavior would only hold till the Rouse time.

In order to do so, we first argue, following Eq. (\ref{e18}), that the
impedance memory kernel $\mu(t)$ --- that connects the forces on and
the velocities of the middle monomer --- scales $\sim
t^{-\frac{2\nu}{1+2\nu}}$. Consider the thought experiment in which we
grab the middle monomer and move it by a small distance $\vec{\delta
r}$ and hold it at its new position; this corresponds to
$\vec{v}(t)=\vec{\delta r}\,\delta(t)$. In time, the ``information''
that the middle monomer has moved to a new position at $t=0$ will
propagate along the backbone of the polymer, and at time $t$, all the
monomers within a backbone distance $n_t\sim t^{\frac{1}{1+2\nu}}$ ---
following the Rouse scaling --- will equilibrate to this new
situation. These $n_t$ equilibrated monomers are however stretched by
an amount $\vec{\delta r}$. With the entropic spring constant of $n$
equilibrated monomers scaling $\sim n^{-2\nu}$, the (restoring) force
the middle monomer would experience at its new position is given by
$\vec{f}(t)\sim n_t^{-2\nu}(-\vec{\delta r})\sim
t^{-\frac{2\nu}{1+2\nu}}(-\vec{\delta r})$ [force $=$ (spring
constant) $\times$ (stretching distance)]. In other words, $\mu(t)\sim
t^{-\frac{2\nu}{1+2\nu}}$. The fluctuation-dissipation theorem then
dictates that
$\langle\vec{f}(t)\cdot\vec{f}(t')\rangle|_{\vec{v}=0}=|\mu(t-t')|\sim(t-t')^{-\frac{2\nu}{1+2\nu}}$
and
$\langle\vec{v}(t)\cdot\vec{v}(t')\rangle|_{\vec{f}=0}=|a(t-t')|\sim(t-t')^{-\frac{2(1+\nu)}{1+2\nu}}$.
Note once again that these relations only hold till the Rouse time; by
the Rouse time the entire polymer is equilibrated to the new position
of the middle monomer, and beyond that time the forces on the middle
monomer are uncorrelated. Finally, the anomalous dynamics of the
middle monomer --- $\langle\Delta r_{N/2}^2\rangle\sim
t^{\frac{2\nu}{1+2\nu}}$ up to the Rouse time and $\sim t$ thereafter
--- is retrieved by integrating
$\langle\vec{v}(t)\cdot\vec{v}(t')\rangle_{\vec{f}=0}$ twice in time.

\begin{figure}[h]
\begin{center}
\includegraphics[angle=270,width=0.55\linewidth]{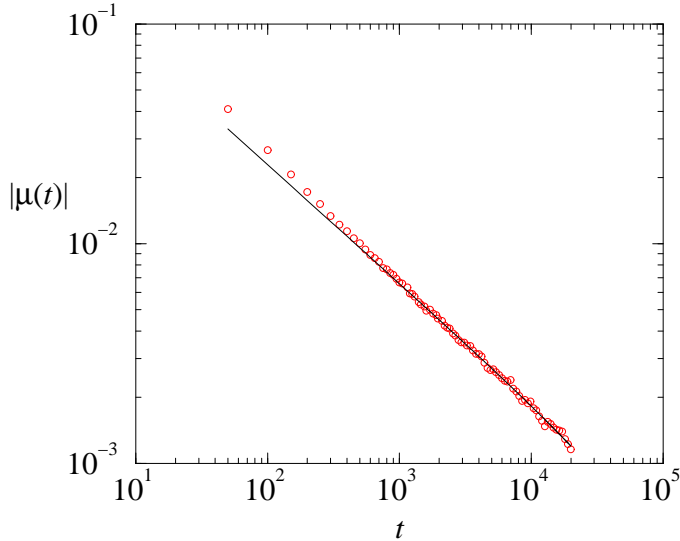}
\end{center}
\caption{The scaling behavior of $|\mu(t)|$ for $N=400$, determined in
the manner described in the text. The solid (black) curve corresponds
to $|\mu(t)|=0.275 t^{-0.54}\exp(-t/\tau_R)$, with
$\tau_R\approx223801$ for $N=400$, obtained from best-fitting the
corresponding data of Fig. \ref{fig2}. \label{fig5}}
\end{figure}

Implementing the above thought experiment into practice and thereby
tracking the restoring force on the middle monomer in order to
determine $\mu(t)$ is a complicated task. Instead, we focus on the
relation $\langle\vec{f}(t)\cdot\vec{f}(t')\rangle_{\vec{v}=0}\sim
t^{-\frac{2\nu}{1+2\nu}}$.  We fix the position of the middle monomer
of a self-avoiding polymer (this corresponds to $\vec{v}=0$) and take
snapshots of it at equal intervals of time. We then take the snapshot
at time $t$, evolve the entire polymer{\it  --- with its middle
monomer free --- over $\Delta t=1$ unit of time for a multiple number
of times}, and obtain the average displacement vector $\vec{u}(t)$ of
the middle monomer. Note that $\Delta t=1$ is small enough such that
the corresponding Eq. (\ref{e19}) --- with $\phi(t)$ and $v(t)$
replaced by $\vec{f}(t)$ and $\vec{v}(t)$ respectively, and
$\langle\vec h(t)\rangle=0$ --- shows that $\vec{u}(t)\propto\vec
f(t)|_{\vec v=0}$. The behavior of
$\langle\vec{u}(t)\cdot\vec{u}(t')\rangle$ in time then yields the
behavior of
$\langle\vec{f}(t)\cdot\vec{f}(t')\rangle|_{\vec{v}=0}=|\mu(t-t')|$.
Determined in the above manner, we confirm $|\mu(t)|\sim
t^{-\frac{2\nu}{1+2\nu}}\exp(-t/\tau_R)$ in Fig. \ref{fig5}, with
$2\nu/(1+2\nu)\approx0.54$, from which, as described in the above
paragraph, the scaling $\langle\Delta r_{N/2}^2\rangle\sim
t^{\frac{2\nu}{1+2\nu}}$ can be derived.

\section{Conclusion\label{sec6}}

To conclude, in this paper we put forward an {\it approximate}
analytical expression (\ref{e3}) for the mode amplitude correlation
functions $X_{pq}(t)$ for long polymers, and corroborate its accuracy
using a lattice-based Monte Carlo simulations of self-avoiding
flexible polymers in three dimensions in the absence of explicit
hydrodynamics.  We report that (a) the mode amplitude correlation
functions are not exactly statistically independent (we do not expect
them to be so anyway); instead, numerically the modes maintain a high
degree of statistical independence, and (b) for higher modes
($p,q\gtrsim6$) there are small deviations from the exponential
behavior (\ref{e3}) at long times. Nevertheless, given that
$X_{pp}(0)$ is a rapidly decaying function of $p$ as per
Eq.~(\ref{e3}), the dominant contribution of the modes to quantities
of interest for the polymer at long time-scales come from the lower
modes. As a result, such quantities can be analytically reconstructed
from Eq.~(\ref{e3}). We demonstrate this by using Eq.~(\ref{e3}) to
derive several scaling properties for self-avoiding
polymers, such as (i) the real-space end-to-end distance, (ii) the
end-to-end vector correlation function, (iii) the correlation function
of the small spatial vector connecting two nearby monomers at the
middle of the polymer, and (iv) the anomalous dynamics of the middle
monomer. Given the content of this paper, we expect that
Eq.~(\ref{e3}) will provide researchers a way forward for analytical
treatment of the properties of self-avoiding polymers.

Importantly, expanding on our recent work on the theory of polymer
translocation, we also demonstrate that the anomalous dynamics of the
middle monomer can be obtained from the forces it experiences, by the
use of the fluctuation-dissipation theorem. Our hope is that in cases
where a polymer's anomalous dynamics exponents are difficult to
identify, use of the fluctuation-dissipation theorem will prove to be
an indispensably useful tool; first such cases were our recent work on
the theory of polymer translocation as well as above.

In experimental situations, the polymer dynamics is often dominated
by hydrodynamic interactions. In future work, we therefore intend to
extend this study to polymer models with explicit hydrodynamics.

\vspace{3mm}
\noindent{\bf Acknowledgments:} D. P. acknowledges ample computer time
on the Dutch national supercomputer facility SARA.

\end{document}